\documentclass[10pt,conference,comsocconf]{IEEEtran}
\IEEEoverridecommandlockouts

\pdfoutput=1
\usepackage[pdftex]{}
\usepackage[acronym]{glossaries}
\usepackage[nohyperlinks,nolist]{acronym}
\usepackage{cite}
\usepackage{amsmath,amssymb,amsfonts}
\usepackage{algorithmic}
\usepackage{graphicx}
\usepackage{textcomp}
\usepackage{xcolor}
\usepackage{orcidlink}
\usepackage{amssymb}
\usepackage{soul}
\usepackage{cancel}
\usepackage{tcolorbox} 
\def\BibTeX{{\rm B\kern-.05em{\sc i\kern-.025em b}\kern-.08em
    T\kern-.1667em\lower.7ex\hbox{E}\kern-.125emX}}
\usepackage{enumitem}
\setlist[itemize]{leftmargin=2.75mm}

\begin{document}
\tcbset{
  colback=gray!20,       
  colframe=black,        
  boxrule=0.5mm,         
  arc=5mm,               
  width=\textwidth      
}

\begin{tcolorbox}
  \begin{center}
    \textbf{THIS IS AN AUTHOR-CREATED POSTPRINT VERSION.} \\[10pt]
    
    \textbf{Disclaimer:} This work has been accepted for publication in \textit{21st IEEE International Conference on Factory Communication Systems (WFCS 2025)}. \\[10pt]
    
    \textbf{Copyright:} \copyright~2025 IEEE. Personal use of this material is permitted. Permission from IEEE must be obtained for all other uses, in any current or future media, including reprinting/republishing this material for advertising or promotional purposes, creating new collective works, for resale or redistribution to servers or lists, or reuse of any copyrighted component of this work in other works. \\[10pt]
    
  \end{center}
\end{tcolorbox}
\newacronym{3GPP}{3GPP}{3rd Generation Partnership Project}

\newacronym{5G}{5G}{5th Generation}
\newacronym{5G-ACIA}{5G-ACIA}{5G Alliance for Connected Industries and Automation}
\newacronym{5GC}{5GC}{5G Core}
\newacronym{5QI}{5QI}{5G QoS Identifier}

\newacronym{6G}{6G}{6th Generation}

\newacronym{AF}{AF}{Application Function}
\newacronym{AI}{AI}{Artificial Intelligence}
\newacronym{AMBR}{AMBR}{Aggregate Maximum Bit Rate}
\newacronym{ARP}{ARP}{Allocation and Retention Priority}
\newacronym{AVB}{AVB}{Audio Video Bridging}

\newacronym{BLER}{BLER}{Block Error Rate}
\newacronym{BS}{BS}{Base Station}

\newacronym{C2C}{C2C}{Controller-to-controller}
\newacronym{C2D}{C2D}{Controller-to-device}
\newacronym{CCDF}{CCDF}{Complementary Cumulative Distribution Function}
\newacronym{CDF}{CDF}{Cumulative Distribution Function}
\newacronym{CF}{CF}{correctionField}
\newacronym{CG}{CG}{Configured Grant}
\newacronym{CNC}{CNC}{Centralized Network Configuration}
\newacronym{CQI}{CQI}{Channel Quality Indicator}
\newacronym{CPE}{CPE}{Customer Premise Equipment}
\newacronym{CPS}{CPS}{Cyber-Physical System}
\newacronym{CRAS}{CRAS}{Connected Robotics and Autonomous Systems}

\newacronym{D2Cmp}{D2Cmp}{Device-to-computer}
\newacronym{DCI}{DCI}{Downlink Control Information}
\newacronym{DG}{DG}{Dynamic Grant}
\newacronym{DL}{DL}{Downlink}
\newacronym{DN}{DN}{Data Network}
\newacronym{DNN}{DNN}{Data Network Name}
\newacronym{DRB}{DRB}{Data Radio Bearer}
\newacronym{DRL}{DRL}{Deep Reinforcement Learning}
\newacronym{DSCP}{DSCP}{Differentiated Services Code Point}
\newacronym{DS-TT}{DS-TT}{Device-side Translator}
\newacronym{D-TDD}{D-TDD}{Dynamic TDD}

\newacronym{E2E}{E2E}{End-to-end}
\newacronym{eMBB}{eMBB}{enhanced Mobile BroadBand}

\newacronym{FD}{FD}{Field Device}
\newacronym{FDD}{FDD}{Frequency Division Duplex}
\newacronym{FIFO}{FIFO}{First-Input First-Output}
\newacronym{FPGA}{FPGA}{Field-Programmable Gate Array}
\newacronym{FRER}{FRER}{Frame Replication and Elimination for Reliability}

\newacronym{GCD}{GCD}{Greatest Common Divisor}
\newacronym{GCL}{GCL}{Gate Control List}
\newacronym{GM}{GM}{Grandmaster}
\newacronym{gNB}{gNB}{next generation Node B}
\newacronym{GNSS}{GNSS}{Global Navigation Satellite System}
\newacronym{PTP}{PTP}{Precision Time Protocol}
\newacronym{GTP}{GTP}{GPRS Tunneling Protocol}

\newacronym{HARQ}{HARQ}{Hybrid Automatic Repeat Request}
\newacronym{HP}{HP}{High Priority}
\newacronym{HMP}{HMP}{High-Medium Priority}

\newacronym{ICI}{ICI}{Inter-Cycle Interference}
\newacronym{IFI}{IFI}{Inter-Flow Interference}
\newacronym{IEEE}{IEEE}{Institute of Electrical and Electronics Engineers}
\newacronym{IoT}{IoT}{Internet of Things}

\newacronym{L2C}{L2C}{Line controller-to-controller}
\newacronym{LCM}{LCM}{Least Common Multiple}
\newacronym{LMP}{LMP}{Low-Medium Priority}
\newacronym{LP}{LP}{Low Priority}
\newacronym{LTE}{LTE}{Long Term Evolution}

\newacronym{MBS}{MBS}{Multicast–Broadcast Services}
\newacronym{MBSFN}{MBSFN}{Multicast-Broadcast Single-Frequency Network}
\newacronym{MB-UPF}{MB-UPF}{Multicast/Broadcast UPF}
\newacronym{MCS}{MCS}{Modulation and Coding Scheme}
\newacronym{MES}{MES}{Manufacturing Execution System}
\newacronym{MNO}{MNO}{Mobile Network Operator}
\newacronym{MTU}{MTU}{Maximum Transmission Unit}
\newacronym{Multi-TRP}{Multi-TRP}{Multiple Transmission and Reception Point}

\newacronym{NIC}{NIC}{Network Interface Card}
\newacronym{NW-TT}{NW-TT}{Network-side Translator}
\newacronym{NR}{NR}{New Radio}
\newacronym{NTP}{NTP}{Network Time Protocol}
\newacronym{NPN}{NPN}{Non-Public Network}

\newacronym{OFDMA}{OFDMA}{Orthogonal Frequency-Division Multiple Access}
\newacronym{OFDM}{OFDM}{Orthogonal Frequency-Division Multiplexing}
\newacronym{O-RAN}{O-RAN}{Open Radio Access Network}


\newacronym{P2P}{P2P}{Peer-to-Peer}
\newacronym{PCP}{PCP}{Priority Code Point}
\newacronym{PDCCH}{PDCCH}{Physical Downlink Control Channel}
\newacronym{PDCP}{PDCP}{Packet Data Convergence Protocol}
\newacronym{PDR}{PDR}{Packet Detection Rule}
\newacronym{PDSCH}{PDSCH}{Physical Downlink Shared Channel}
\newacronym{PDU}{PDU}{Packet Data Unit}
\newacronym{PLC}{PLC}{Programmable Logic Controller}
\newacronym{PPS}{PPS}{Pulse Per Second}
\newacronym{PRACH}{PRACH}{Physical Random Access Channel}
\newacronym{PTP}{PTP}{Precision Time Protocol}
\newacronym{PUCCH}{PUCCH}{Physical Uplink Control Channel}
\newacronym{PUSCH}{PUSCH}{Physical Uplink Shared Channel}

\newacronym{QAM}{QAM}{Quadrature Amplitude Modulation}
\newacronym{QoE}{QoE}{Quality of Experience}
\newacronym{QoS}{QoS}{Quality of Service}
\newacronym{QFI}{QFI}{QoS Flow ID}

\newacronym{RAN}{RAN}{Radio Access Network}
\newacronym{RB}{RB}{Resource Block}
\newacronym{RLC}{RLC}{Radio Link Control}
\newacronym{RRC}{RRC}{Radio Resource Control}
\newacronym{RTC1}{RTC1}{Real Time Class 1}

\newacronym{SCADA}{SCADA}{Supervisory Control and Data Acquisition}
\newacronym{SCS}{SCS}{Sub-Carrier Spacing}
\newacronym{SDR}{SDR}{Software Defined Radio}
\newacronym{SFP}{SFP}{Small Form-factor Pluggable}
\newacronym{SMF}{SMF}{Session Management Function}

\newacronym{SINR}{SINR}{Signal-to-Interference-plus-Noise Ratio}
\newacronym{SPS}{SPS}{Semi-Persistent Scheduling}
\newacronym{SST}{SST}{Slice/Service Type}
\newacronym{S-NSSAI}{S-NSSAI}{Single Network Slice Selection Assistance Information}

\newacronym{TAS}{TAS}{Time-Aware Shaping}
\newacronym{TC}{TC}{Transparent Clock}
\newacronym{TDD}{TDD}{Time Division Duplex}
\newacronym{TDMA}{TDMA}{Time-Division Multiple Access}
\newacronym{TSC}{TSC}{Time-Sensitive Communication}
\newacronym{TSN}{TSN}{Time-Sensitive Networking}
\newacronym{TTI}{TTI}{Transmission Time Interval}

\newacronym{UDP}{UDP}{User Datagram Protocol}
\newacronym{UE}{UE}{User Equipment}
\newacronym{UFTP}{UFTP}{UDP-based File Transfer Protocol}
\newacronym{UL}{UL}{Uplink}
\newacronym{UPF}{UPF}{User Plane Function}
\newacronym{URLLC}{URLLC}{Ultra-Reliable and Low-Latency Communications}
\newacronym{uRLLC}{uRLLC}{ultra-Reliable and Low-Latency Communications}

\newacronym{VLAN}{VLAN}{Virtual Local Area Network}
\newacronym{VNF}{VNF}{Virtualized Network Function}
\newacronym{VNI}{VNI}{Virtual Network Identifier}
\newacronym{VTEP}{VTEP}{VxLAN Tunnel End Point}
\newacronym{VxLAN}{VxLAN}{Virtual Extensible LAN}
\newacronym{vBBU}{vBBU}{virtual Baseband Unit}

\title{Empirical Analysis of the Impact of 5G Jitter on Time-Aware Shaper Scheduling in a \\5G-TSN Network\\

\thanks{This work has been financially supported by the Ministry for Digital Transformation and of Civil Service of the Spanish Government through TSI-063000-2021-28 (6G-CHRONOS) project, and by the European Union through the Recovery, Transformation and Resilience Plan - NextGenerationEU. Additionally, this publication is part of grant PID2022-137329OB-C43 funded by MICIU/AEI/ 10.13039/501100011033 and part of FPU Grant 21/04225 funded by the Spanish Ministry of Universities.}
}

\author{\IEEEauthorblockN{Pablo Rodriguez-Martin\IEEEauthorrefmark{1}\IEEEauthorrefmark{2}, Oscar Adamuz-Hinojosa\IEEEauthorrefmark{1}\IEEEauthorrefmark{2}, Pablo Muñoz\IEEEauthorrefmark{1}\IEEEauthorrefmark{2}, \\ Julia Caleya-Sanchez\IEEEauthorrefmark{1}\IEEEauthorrefmark{2}, Jorge Navarro-Ortiz\IEEEauthorrefmark{1}\IEEEauthorrefmark{2},
Pablo Ameigeiras\IEEEauthorrefmark{1}\IEEEauthorrefmark{2}}

\IEEEauthorblockA{\IEEEauthorrefmark{1}Department of Signal Theory, Telematics and Communications, University of Granada.}
\IEEEauthorblockA{\IEEEauthorrefmark{2}Research Center on Information and Communication Technologies, University of Granada.}
\IEEEauthorblockA{Email: \{pablorodrimar, oadamuz, pabloml, jcaleyas, jorgenavarro, pameigeiras\}@ugr.es\IEEEauthorrefmark{1}\IEEEauthorrefmark{2} }
}


\maketitle

\begin{abstract}
Deterministic communications are essential for industrial automation, ensuring strict latency requirements and minimal jitter in packet transmission. Modern production lines, 
specializing
in robotics, require higher flexibility and mobility, which drives the integration of Time-Sensitive Networking (TSN) and 5G networks in Industry 4.0. TSN achieves deterministic communications by using mechanisms such as the IEEE 802.1Qbv Time-Aware Shaper (TAS), which schedules packet transmissions within precise cycles, thereby reducing latency, jitter, and congestion. 5G networks complement TSN by providing wireless mobility and supporting ultra-Reliable Low-Latency Communications. However, 5G channel effects such as fast fading, interference, and network-induced latency and jitter can disrupt TSN traffic, potentially compromising deterministic scheduling and performance. This paper presents an empirical analysis of 5G network latency and jitter on IEEE 802.1Qbv performance in a 5G-TSN network. We evaluate the impact of 5G integration on TSN’s deterministic scheduling through a testbed combining IEEE 802.1Qbv-enabled switches, TSN translators, and a commercial 5G system. Our results show that, with proper TAS configuration in the TSN switch aligned with the 5G system, jitter can be mitigated, maintaining deterministic performance.

\end{abstract}


\section{Introduction}
The advent of \gls{5G} networks and the emerging vision for \gls{6G} networks are driven by the need to support a broad range of applications, including \gls{CRAS}\cite{Saad20}. These systems—encompassing autonomous and collaborative robots, drones, and other intelligent agents—have the potential to revolutionize multiple industry verticals, with industrial automation standing out as a key beneficiary. The integration of cutting-edge technologies such as the \gls{IoT}, \gls{CPS}, cloud computing, and \gls{AI} is expected to elevate industrial automation to unprecedented levels of efficiency and intelligence.

A fundamental requirement for industrial automation applications is the ability to provide communication services that meet stringent constraints in terms of data rate, reliability, and latency \cite{IIC2019}. This necessitates network architectures capable of providing efficient, low-latency, highly reliable, and deterministic communications. To address these needs, the 
IEEE
has developed the \gls{TSN} standards, which introduce key capabilities such as time synchronization, stream reservation, traffic shaping, scheduling, preemption, traffic classification, and seamless redundancy. These features are essential for ensuring the deterministic performance required in industrial automation environments.

While \gls{TSN} standards ensure determinism and reliability in wired networks, \gls{5G}/\gls{6G} systems extend these capabilities by enabling mobility, wide-area coverage, and scalability. Integrating \gls{5G}/\gls{6G} systems with \gls{TSN}~\cite{5GACIA-whitepaperI} has emerged as a promising solution to meet the stringent communication demands of \gls{CRAS} and industrial automation. In this setup, production lines in a factory connect wirelessly to the \gls{5G}/\gls{6G} system, which then interfaces with the enterprise edge-cloud through \gls{TSN} switches. However, this integration presents significant challenges due to fundamental differences between wired \gls{TSN} networks and wireless \gls{5G}/\gls{6G} systems. A key challenge lies in \emph{incorporating \gls{TSN}'s IEEE 802.1Qbv standard, which defines a \gls{TAS} for deterministic packet scheduling}. The introduction of \gls{5G}/\gls{6G} systems into a \gls{TSN}-based industrial network introduces additional latency and jitter compared to traditional \gls{TSN} switches and wired links. These impairments stem from the dynamic and uncertain nature of wireless transmission and the time-varying processing delays within \gls{5G}/\gls{6G} network nodes. The jitter introduced by \gls{5G}/\gls{6G} may cause packets to not be transmitted according to the \gls{TAS} schedules, resulting in potential failures in industrial processes. Thus, a dejittering mechanism based on \gls{TAS} is needed to maintain the deterministic behavior in the network.

\subsection{Related work}
Currently, the research on integration between \gls{5G} and \gls{TSN} networks is in its early stages. The works in \cite{Sat23}\cite{Sas24} provide a comprehensive analysis of the current and future research directions on \gls{5G}-\gls{TSN} integration. From an architectural perspective, it is well established that the \gls{5G} system behaves as a \gls{TSN} logical switch, as discussed in \cite{Ros22}\cite{Lar20}. Several works address time synchronization \cite{Rod22} and \gls{5G}-\gls{TSN} QoS mapping \cite{Deb23} as key functions for this logical switch model. However, there are still some open issues to be addressed. A crucial issue is related to the non-deterministic behavior of the \gls{5G} system and its impact on the \gls{TSN} scheduled traffic. This challenge has been addressed in \cite{Fon24}\cite{Wang24}, where evaluations are carried out by simulation tools. In relation to this, there is also a lack of 
functional
testbeds to conduct tests under realistic conditions. Some preliminary results on this are presented in \cite{Aij24}, revealing a significant dependence on the native capabilities of the \gls{5G} system and a need for a global schedule in the integrated network. Thus, further research is required to analyze the impact of the \gls{5G} capabilities and constraints.

\subsection{Contributions}
In this work, we present an empirical analysis of the impact of \gls{5G} network latency and jitter on the performance of IEEE 802.1Qbv in an integrated \gls{5G}-\gls{TSN} network. Our key contributions are as follows:
\begin{itemize}
    \item[C1] We analyze the \gls{DL} transmission in a \gls{5G}-\gls{TSN} network and evaluate how the integration of the \gls{5G} system affects the deterministic scheduling of IEEE 802.1Qbv.
    \item[C2] We develop an experimental testbed integrating \gls{TSN} and \gls{5G} technologies, including IEEE 802.1Qbv-enabled switches, \gls{TSN} translators, and a commercial \gls{5G} system.
    \item[C3] We identify the key configuration parameters required to optimize IEEE 802.1Qbv integration in a \gls{5G}-\gls{TSN} environment and analyze the associated constraints.
\end{itemize}

Our results demonstrate that, with an appropriate configuration of the \gls{TAS} schedule in the \gls{TSN} switches, it is possible to mitigate the \gls{5G} jitter and maintain deterministic performance. 

The paper is structured as follows. Section~\ref{sec:Background} provides background on industrial \gls{5G}-\gls{TSN} networks. Section~\ref{sec:SystemModel} introduces the system model, while Section~\ref{sec:jitter} analyzes the impact of \gls{5G} jitter on \gls{TAS} scheduling. Section~\ref{sec:Testbed} describes the testbed and experimental setup. In Section~\ref{sec:Results}, we present the performance results. Finally, Section~\ref{sec:Conclusions} summarizes the key conclusions and outlines directions for future work.

\section{Background on 5G-TSN Industrial Networks}\label{sec:Background}
In this section, we provide background on the network architecture, device synchronization, deterministic traffic shaping, and the considered industrial applications.

\subsection{Network Segments for Industry Automation}

\begin{figure}[t!]
    \centering
   \includegraphics[scale=0.316]{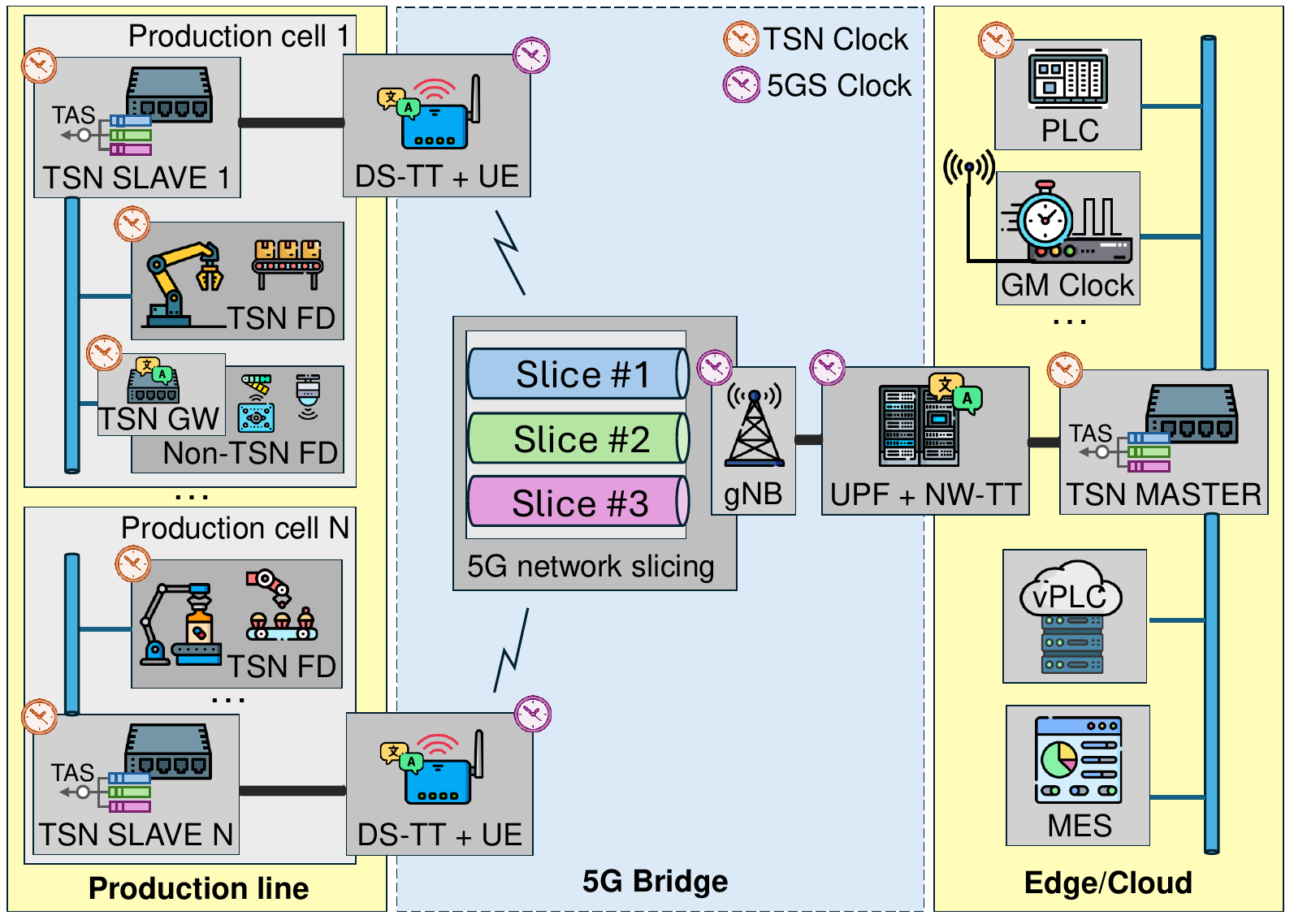}
    \caption{5G-TSN network architecture in an Industry 4.0 factory.}
   \label{fig:architecture}
\end{figure}

As depicted in Fig. \ref{fig:architecture}, three connectivity segments may be identified in a \gls{5G}-\gls{TSN}-based industrial network~\cite{5GACIA-whitepaperI}:

\begin{itemize}
     \item \textit{\textbf{Edge/Cloud Room}}:  serves as a centralized management layer, hosting control functions such as those traditionally executed by \glspl{PLC}. These can be deployed on dedicated hardware or general-purpose servers. 
     Additionally, management tasks—including monitoring, data collection, and analytics—are handled by systems like \gls{MES}. This segment also incorporates a server providing the \gls{TSN} \gls{GM} clock reference to be distributed, typically derived from \gls{GNSS}.

    \item \textit{\textbf{5G System}}: following \gls{3GPP} TS 23.501 (v19.0.0)~\cite{3gpp_ts_23_501_v19_0_0}, the \gls{5G} system 
    integrates into the \gls{TSN} network as one or more virtual \gls{TSN} switches, with \glspl{UPF} and \glspl{UE} as endpoints. The \gls{UE} connects to the \gls{UPF} via the \gls{gNB}, which provides wireless connectivity. The \gls{TSN} translators, i.e., \gls{NW-TT} and \gls{DS-TT}, integrated within the \gls{UPF} and \gls{UE}, respectively; handle wireless \gls{5G} network timing correction functionalities. This work focuses solely on the data plane, excluding control functions, and considers a single \gls{UE} per production line, which provides wireless \gls{5G} connectivity between production lines and enables access to the edge/cloud room.

    \item \textit{\textbf{Production Line}}: it includes multiple \glspl{FD} and local \glspl{PLC} for distributed control. \glspl{FD} report status and measurements to centralized \glspl{PLC}, enabling hierarchical decision-making. The production line is divided into cells, each connected to a \gls{TSN} slave switch that receives the clock signal from the \gls{TSN} master via the \gls{5G} system and redistributes it to the \glspl{FD} in its production line. 

    
\end{itemize}


Focusing on traffic management, traffic from various industrial applications may be isolated in the \gls{5G} system into distinct \gls{QoS} flows. Each flow is treated differently in terms of resource allocation, based on its specific requirements for latency, reliability, bandwidth, and priority, as defined in 3GPP TS 23.501 (v19.0.0). 
To harmonize \gls{QoS} management between \gls{TSN} and \gls{5G} system, packet filters at the \gls{UE} and \gls{UPF} map Ethernet packets to specific \gls{5G} \gls{QoS} flows based on their \gls{PCP} values. In \gls{UL}, the \gls{UE} applies packet filtering to classify outgoing packets before they enter the \gls{5G} network. In \gls{DL}, the \gls{UPF} performs this filtering to ensure packets are assigned to the appropriate \gls{5G} \gls{QoS} flow before reaching the \gls{UE}. The \gls{PCP} is a field within the \gls{VLAN} tag used to classify Ethernet frames into different priority levels. Additionally, one or more \gls{5G} \gls{QoS} flows can be mapped to a specific network slice, enabling the reservation of a dedicated amount of radio resources for that slice, and thus for those \gls{5G} \gls{QoS} flows.




\subsection{Synchronization in 5G-TSN Network}\label{sec:sync}
Time synchronization is essential in a \gls{5G}-\gls{TSN} network to ensure the deterministic execution of industrial automation processes. In this context, a \gls{TSN} network typically includes a \gls{TSN} master switch that distributes the \gls{GM} clock reference via \gls{PTP} messages to one or more \gls{TSN} slave switches in the production lines. These \gls{TSN} slaves compute the clock offset from the \gls{TSN} master switch to adjust their clocks accordingly. We assume the \gls{5G} system operates as a \gls{PTP} \gls{TC} \cite{StandadsIEEE1588_2019}, meaning that the \gls{5G} bridge forwards synchronization messages without modifying the timestamps but correcting delay between \gls{TSN} translators. 

Discrepancies in the clocks of different devices within the \gls{5G}-\gls{TSN} network can occur, preventing the devices from updating their clocks accurately. This leads to clock drifts, typically in the order of hundreds of nanoseconds. \gls{3GPP} TS 22.104 \cite{3gpp_ts22104} specifies a maximum delay of 900 ns must be guaranteed in the industrial network for \gls{5G} systems. However, the synchronization error is several orders of magnitude smaller than the \gls{5G} jitter, which is in the order of milliseconds. Therefore, we assume the synchronization error is negligible.

\subsection{IEEE 802.1Qbv Time-Aware Shaper}
IEEE 802.1Qbv is a \gls{TSN} standard that defines the \gls{TAS} mechanism, enabling time-aware scheduling of Layer 2 frames at each \gls{TSN} switch’s egress port while accounting for different \gls{QoS} levels \cite{Oge2020,Oliver2018,Walrand2023}. \gls{TAS} operates based on the \gls{PCP} field in the IEEE 802.1Q header, where frames are assigned a 3-bit priority value, i.e. from 0 to 7, and mapped to one of eight \gls{FIFO} queues. These queues, located at each switch’s egress port, correspond to distinct traffic classes, ensuring differentiated \gls{QoS} enforcement.

To regulate frame transmission from these queues, each egress port is associated with a \gls{GCL}, which determines the eligibility of each queue for transmission. Each queue has its own gate, and traffic scheduling is organized into cyclical
\emph{transmission windows},
during which one or more gates open to allow transmission from the respective queues. Within each transmission window, if multiple gates are open simultaneously, frames from the highest-priority queue are transmitted first, followed by frames from the next highest-priority queue, and so on, until all eligible queues have been served. For \gls{TAS}, we assume one gate at a time.

\subsection{Industrial applications}
In industrial networks, most of the traffic is delay-critical, as defined in \cite{5GACIA-whitepaperI}. It can be classified as \emph{isochronous} or \emph{cyclic}.

Isochronous applications require strictly periodic and coordinated packet transmissions, with delays shorter than their execution period, a.k.a. \emph{application cycle}. These high-priority applications, common in motion control, automotive systems, \gls{AI} vision, and \gls{AVB}; rely on synchronized clocks across all \gls{TSN} switches. Each switch and its \gls{TAS} scheduling must be time-aligned to prevent delays, ensuring packets reach the processing unit (e.g., \gls{PLC} at the edge) on time. In contrast, cyclic applications also follow a periodic transmission pattern but are less stringent about synchronization among \gls{TSN} switches. These applications can tolerate bounded jitter, and times may vary depending on when the industrial process is initiated. Examples include industrial sensor polling, where data is periodically collected and processed. This work focuses on general cyclic applications.

\section{System Model}\label{sec:SystemModel}
In this section, we first present the network model, which defines the fundamental structure of the industrial network under consideration. Next, we introduce the traffic model, describing the characteristics of the considered industrial data flows. Finally, we outline the \gls{TAS} model, detailing its role within the industrial network.

\subsection{Network Model} \label{sec:network_model}
\begin{figure}[b!]
\includegraphics[width=\columnwidth]{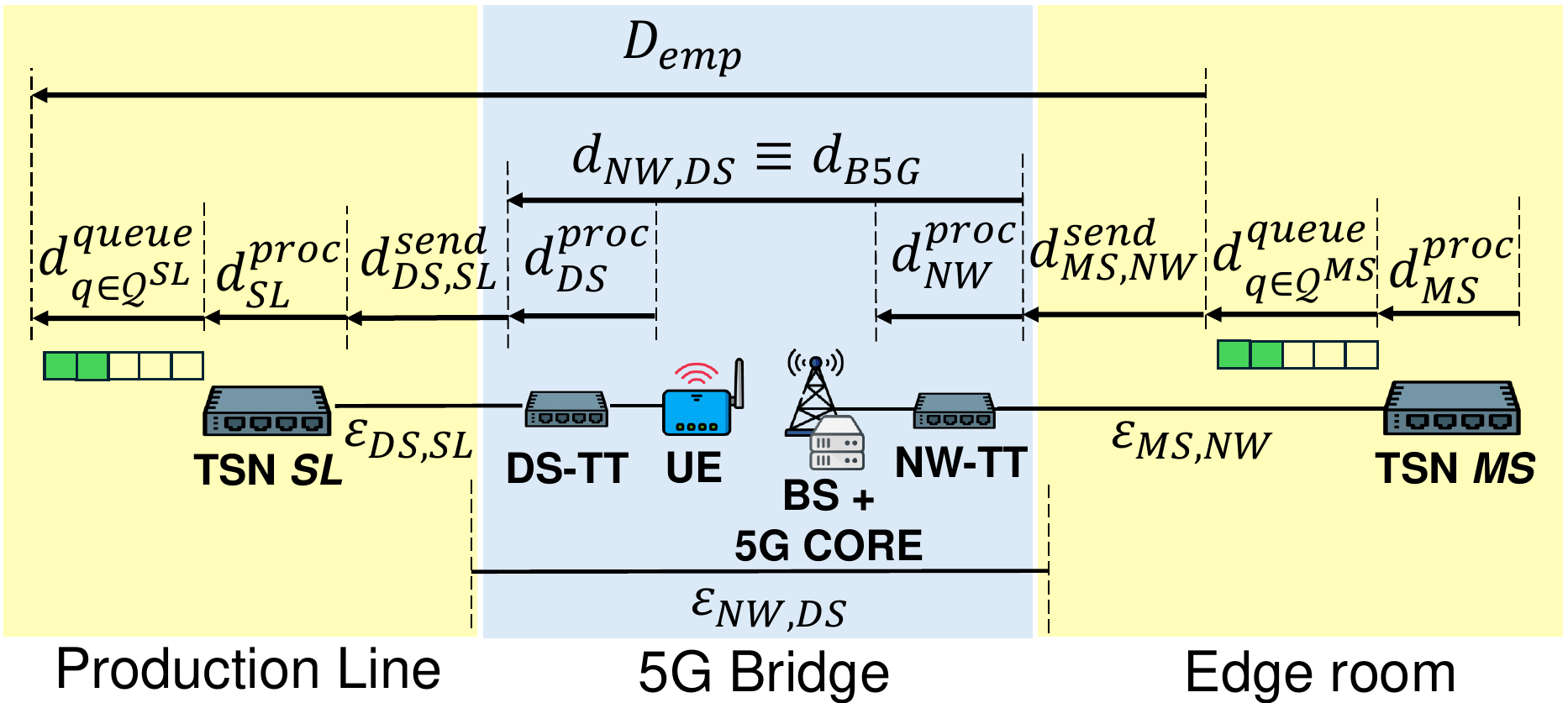}
\caption{5G-TSN network elements and associated delays in a \gls{DL} transmission.}
\label{fig:network_model}
\end{figure}

We consider an industrial network $\mathcal{V}$ comprising two \gls{TSN} switches: a \gls{TSN} master $\text{MS}$ and a \gls{TSN} slave $\text{SL}$, where $\text{MS}, \text{SL} \in \mathcal{V}$, as depicted in Fig. \ref{fig:network_model}. The processing delay of these \gls{TSN} switches, i.e., $d_{\text{MS}}^{proc}$ and $d_{\text{SL}}^{proc}$, represent the time required to internally transfer a packet to an output port's queue. 
Additionally, the output ports of \gls{TSN} switches $\text{MS}$ and $\text{SL}$ operate with a macrotick of $m$ seconds.

The transmission between \gls{TSN} switches $\text{MS}$ and $\text{SL}$ is carried out by several links and devices. The equipment supporting this packet transmission consists of two \gls{TSN} translators, i.e., \gls{NW-TT} and \gls{DS-TT}, as well as the \gls{UE}, the \gls{gNB} and the \gls{5G} core, thus shaping our \gls{5G} bridge.
Both \gls{NW-TT} and \gls{DS-TT} are denoted as $\text{NW}$ and $\text{DS}$, respectively, with $\text{NW}, \text{DS} \in \mathcal{V}$. Both \gls{TSN} translators have processing delays denoted as $d_{\text{NW}}^{proc}$ and $d_{\text{DS}}^{proc}$. Every link is denoted with $\varepsilon$, i.e. $\varepsilon_{\text{MS},\text{NW}}, \varepsilon_{\text{DS},\text{SL}} \in \mathcal{E}$. 

 Since links are bidirectional, we distinguish between the \gls{UL} and \gls{DL} by swapping the subscript order, i.e., $\varepsilon_{\text{NW},\text{MS}}, \varepsilon_{\text{SL},\text{DS}} \in \mathcal{E}$. The bandwidth of each link is given by $b$, i.e., $b_{\text{MS},\text{NW}}$ and $b_{\text{DS},\text{SL}}$, respectively. The bandwidth is established by the ports' capacities. In terms of delay, the \emph{propagation delays} are determined by $d_{\text{MS},\text{NW}}^{prop}$ and $d_{\text{DS},\text{SL}}^{prop}$. The total \emph{sending delay} is equivalent to adding the \emph{transmission delay} to the propagation delay, i.e., $d_{\text{MS},\text{NW}}^{send} = d_{\text{MS},\text{NW}}^{prop} + l_x/b_{\text{MS},\text{NW}}$, where $l_x$ is the length of the bit string to be transmitted. Similarly, $d_{\text{DS},\text{SL}}^{send} = d_{\text{DS},\text{SL}}^{prop} + l_x/b_{\text{DS},\text{SL}}$.

The \gls{5G} bridge, functioning as a \gls{TSN} non-time-aware device, forwards packets between virtual ports, i.e., \gls{NW-TT} and \gls{DS-TT}; without implementing any \gls{TAS} configuration. It comprises both \gls{TSN} translators and the \gls{5G} system wireless link $\varepsilon_{\text{NW}, \text{DS}} \in \mathcal{E}$, which defines the inner connection between \gls{5G} system elements.

For wired connections, the behavior is typically symmetric between \gls{UL} and \gls{DL} in terms of data rate and latency, e.g., $b_{\text{DS},\text{SL}} = b_{\text{SL},\text{DS}}$ and $d_{\text{DS},\text{SL}}^{send} = d_{\text{SL},\text{DS}}^{send}$. In contrast, for wireless connections in the \gls{5G} bridge, both data rate and latency must be measured separately for \gls{UL} and \gls{DL} due to inherent capacity asymmetry, i.e., $b_{\text{NW},\text{DS}} \neq b_{\text{DS},\text{NW}}$ and $d_{\text{NW},\text{DS}} \neq d_{\text{DS},\text{NW}}$. For simplicity, we denote this \gls{5G} bridge delay as $d_{\text{B5G}}$. We also denote $b_{\text{MS},\text{SL}}$ as the composed capacity of all wired and wireless links in series from MS to SL, likely equal to the bottleneck capacity of \gls{5G} bridge in \gls{DL}, $b_{\text{NW},\text{DS}}$.

Furthermore, we define the queuing delay for the queue $q \in \mathcal{Q}^x$ at the output port of the \gls{TSN} switches, where $x=\left\{\text{MS},\text{SL}\right\}$ denotes a specific \gls{TSN} switch. The queuing delay at the \gls{TSN} switch $\text{MS}$ for queue $q$ is denoted as  $d_{q\in\mathcal{Q}^\text{MS}}^{queue}$, and at the \gls{TSN} switch $\text{SL}$ for queue $q$ as  $d_{q\in\mathcal{Q}^\text{SL}}^{queue}$.

\subsection{Traffic Model} \label{sec:traffic_model}
In this work, we focus on the \gls{DL} traffic, although the formulation presented hereafter can also be extrapolated to the \gls{UL} direction. We consider two types of traffic streams: a \emph{delay-critical} stream, denoted as $\text{DC} \in \mathcal{S}$, and a \emph{best-effort} stream, denoted as $\text{BE} \in \mathcal{S}$, where $\mathcal{S}$ is the set of streams. 

For the delay-critical stream $\text{DC}$, we assume a set of $N_{\text{DC}}$ packets that are transmitted in a burst, with a fixed \emph{application cycle} $T_{\text{DC}}^{app}$ between transmissions. 
The packet size $l_{\text{DC}}$ for traffic stream $\text{DC}$ is considered constant. Additionally, we define the packet delay budget $D_{\text{DC}}$ for the delay-critical stream $\text{DC}$ within the industrial network requirements. For the best-effort stream $\text{BE}$, we assume that packets are transmitted at a constant data rate $R_{\text{BE}}$. The packet size $l_{\text{BE}}$ for $\text{BE}$ stream is also constant. Additionally, no packet delay budget is defined for the best-effort stream $\text{BE}$.

We consider a \gls{5G} network slice associated to a \gls{PCP} for each kind of traffic. Thus, \gls{uRLLC} slice is reserved for DC traffic while \gls{eMBB} is for BE traffic.

\subsection{TAS model}\label{sec:TAS_model} 

The \textit{network cycle} \cite{Lin22} is a fixed and recurring time interval during which 
the scheduled transmissions from the considered output port queues are organized and executed according to a predefined \gls{GCL} schedule. Each network cycle consists of multiple \textit{transmission windows}, ensuring deterministic communication by assigning exclusive transmission opportunities to different traffic classes. The duration of the network cycle corresponds to the \gls{GCD} of all \textit{application cycles} involved, given by $T_C = \operatorname{GCD}(T_s^{app})$ $\forall s \in  \mathcal{S}$. 
In our study, since only one \textit{delay-critical} stream $\text{DC}$ is constrained by an application cycle of $T_{\text{DC}}^{app}$, it results in $T_C = T_{\text{DC}}^{app}$.

We assume the \gls{GCL} enforces exclusive stream transmission, allowing only one queue to transmit at a \textit{transmission window}. Focusing on the \textit{delay-critical} stream $\text{DC}$, we examine its transmission from \gls{TSN} switch $\text{MS}$ to \gls{TSN} switch $\text{SL}$. Specifically, we consider the corresponding queue $q \in \mathcal{Q}^{\text{MS}}$ at the output port of \gls{TSN} switch $\text{MS}$. The \gls{GCL} for this port must be configured based on the \textit{network cycle} $T_C$, as outlined in Eq.~\eqref{eqn:cgl_open}. The binary variable $G_{\text{MS},\text{SL},q\to\text{DC}}(t)$ takes the value 1 when the gate is open and 0 otherwise. The gate opens periodically during each \textit{network cycle} of duration $T_C$. Focusing on the first network cycle, i.e., $n=0$, the variables $t_{1,\text{DC}}^{\text{MS},\text{SL}}$ and $t_{2,\text{DC}}^{\text{MS},\text{SL}}$ define the time instants at the \textit{transmission window} for the traffic stream $\text{DC}$, thereby establishing when the gate is open or closed, respectively. 
\begin{equation}
{\small
G_{\text{MS},\text{SL}_{q\to\text{DC}}}(t) =
\begin{cases} 
    1, &  nT_C+t_{1,\text{DC}}^{\text{MS},\text{SL}} < t \leq  nT_C+t_{2,\text{DC}}^{\text{MS},\text{SL}},\\
     &  \; \forall n  \in \mathbb{N} \cup \{0\} \\
    0, &  otherwise
\end{cases} }
\label{eqn:cgl_open}
\end{equation}

Based on that, we can define the duration of the \textit{transmission window} as $w^{\text{MS},\text{SL}}_{\text{DC}} = t_{2,\text{DC}}^{\text{MS},\text{SL}} - t_{1,\text{DC}}^{\text{MS},\text{SL}}$. The duration  $w^{\text{MS},\text{SL}}_{\text{DC}}$ of the \textit{transmission window}  for the traffic stream $\text{DC}$ must be at least as large as the transmission time required for the burst of $N_{\text{DC}}$ packets and lower than the 
\textit{network cycle} $T_{C}$
as defined in Eq.~\eqref{eqn:window_bounds}.  The lower size bound is determined by multiplying the number of packets $N_{\text{DC}}$ by their size $l_{DC}$, and then dividing by the link's bandwidth $b_{\text{MS},\text{SL}}$. 
\begin{equation}
N_{\text{DC}}\frac{l_{\text{DC}}}{b_{\text{MS},\text{SL}}} \leq w^{\text{MS},\text{SL}}_{\text{DC}} < T_{C}
\label{eqn:window_bounds}
\end{equation}

Note that, if the previous condition is not met, the following consequences may arise. If the \textit{transmission window} is shorter than its lower bound, i.e.,  $w^{\text{MS},\text{SL}}_{\text{DC}} < N_{\text{DC}} l_{\text{DC}}/b_{\text{MS},\text{SL}}$, it means insufficient time to transmit the $N_{\text{DC}}$ packets of the traffic flow $\text{DC}$. As a result, the packets that remain queued are transmitted with an extra delay that is a multiple of the \emph{network cycle} duration $T_{C}$, since they would accumulate and be transmitted in the subsequent network cycles, leading to a packet queueing effect. On the other hand, if the \textit{transmission window} exceeds the upper bound, i.e., $w^{\text{MS},\text{SL}}_{\text{DC}} > T_{C}$, the scheduling would not be approachable as it takes place during the whole network cycle and no other streams can be transmitted. Moreover, if $w^{\text{MS},\text{SL}}_{\text{DC}}$ is not well fitted, the elapsed time between the end of the transmission of the last packet and the end of the network cycle would be wasted.

In our study, the remaining time until the completion of the \emph{network cycle} $T_{C}$ is used to transmit data from the \emph{best-effort} stream $\text{BE} \in \mathcal{S}$, followed by a guard band to avoid frame collisions. This means $T_C = w^{\text{MS},\text{SL}}_{\text{DC}}  + w^{\text{MS},\text{SL}}_{\text{BE}}  +t_{GB}$, where $w^{\text{MS},\text{SL}}_{\text{BE}} $ is the \textit{transmission window} for the \emph{best-effort} stream $\text{BE}$ and $t_{GB}$ the guard band time.

After establishing the \gls{TAS} model, a mathematical analysis of its behavior in a \gls{5G}-\gls{TSN} network is presented below.


\section{Analysis of 5G jitter on TAS scheduling}\label{sec:jitter}

The traffic streams $\text{DC},\text{BE} \in \mathcal{S}$ will follow a path of network nodes until reaching its destination as illustrated in Fig. \ref{fig:network_model}. This path is defined as $\mathcal{P}_{nodes} \equiv \left\{\text{MS},\text{NW},\text{DS}, \text{SL}\right\}$, where $\text{MS},\text{NW},\text{DS}, \text{SL} \in \mathcal{V}$. Additionally, the list of links that connect these nodes is defined as $\mathcal{P}_{links} \equiv \left\{\varepsilon_{\text{MS},\text{NW}},\varepsilon_{\text{NW},\text{DS}},\varepsilon_{\text{DS},\text{SL}}\right\}$, where  $\varepsilon_{\text{MS},\text{NW}},\varepsilon_{\text{NW},\text{DS}},\varepsilon_{\text{DS},\text{SL}} \in \mathcal{E}$. The total packet transmission delay $D_{tot}$ for a stream $s\in \mathcal{S}$ following this path is described by Eq.~\ref{eqn:e2e_delay}.

\begin{equation}
    D_{tot}=\hspace{-0.5cm}\sum_{\forall v \in \left\{\text{MS},\text{SL}\right\}}\hspace{-0.25cm}\left(d_{q\in\mathcal{Q}^{v}}^{queue}+d_{v}^{proc}\right) +d_{\text{MS},\text{NW}}^{send}+d_{\text{B5G}}+d_{\text{DS},\text{SL}}^{send}
\label{eqn:e2e_delay}
\end{equation}

In our work, we focus on an analysis of the \gls{5G} jitter on \gls{TAS} scheduling. The jitter is defined as the variation of the packet transmission delay in the \gls{5G} network due to channel effects as fast-fading. To properly capture the impact of the  \gls{5G} jitter on \gls{TAS} scheduling, we empirically measure the packet transmission delay $D_{emp}$ since the \gls{TSN} $\text{MS}$ sends the packet until this packet is departed from the \gls{TSN} $\text{SL}$. Based on that, the packet transmission delay $D_{emp}$ can be defined as Eq.~\eqref{eqn:empirical_delay} shows. Note that $K=d_{\text{MS},\text{NW}}^{send}+d_{\text{DS},\text{SL}}^{send}+d_{\text{SL}}^{proc}$ represents the sum of those delay terms that are constant, being the remaining terms variable for each packet. The term $d_{\text{B5G}}$ includes the \gls{5G} jitter. Concerning $d_{q\in\mathcal{Q}^{\text{SL}}}^{queue}$, its variability depends on factors such as: (a) the time the packet enters the corresponding queue, (b) the current queue occupation, and (c) if the \textit{transmission window}—and thus, the gate—is open or not for such queue. 
\begin{equation}
\begin{split}
    D_{emp}&= d_{\text{MS},\text{NW}}^{send} + d_{\text{B5G}} + d_{\text{DS},\text{SL}}^{send} +d_{\text{SL}}^{proc} + d_{q\in\mathcal{Q}^{\text{SL}}}^{queue} =\\
    &= K + d_{\text{B5G}} + d_{q\in\mathcal{Q}^{\text{SL}}}^{queue} 
    \end{split}
\label{eqn:empirical_delay}
\end{equation}

To reduce the variability of the delay term $d_{q\in\mathcal{Q}^{\text{SL}}}^{queue}$, the sequence of \gls{GCL} events needs to be coordinated among the \gls{TSN} switches $\text{MS},\text{SL}\in\mathcal{V}$. Specifically, the receiving \gls{TSN} switch $\text{SL}$ must apply an offset $\delta$ to the beginning of the \textit{network cycle} concerning the transmitter \gls{TSN} switch $\text{MS}$. The use of this offset aims to ensure the received packets from the \gls{TSN} switch $\text{MS}$ arrive at the \gls{TSN} $\text{SL}$ before a \emph{transmission window} starts. To achieve this, the offset $\delta$ must have the lower bound defined in Eq.~\eqref{eqn:upperbound_offset}. Specifically, this lower bound is the sum of all the packet propagation delays from this packet leaves the \gls{TSN} $\text{MS}$ until it reaches the \gls{TSN} $\text{SL}$, including the processing delay $d_{\text{SL}}^{proc} $ required by the  \gls{TSN} $\text{SL}$ to put this packet into the corresponding queue. Note that the delays $d_{\text{MS},\text{NW}}^{send}$, $d_{\text{DS},\text{SL}}^{send}$ and $d_{\text{SL}}^{proc}$ are constant whereas $d_{\text{B5G}}$ is variable due to the fluctuating packet transmission time in the \gls{5G} network, caused by wireless channel effects such as fast fading. For such reason, Eq.~\eqref{eqn:upperbound_offset} considers the maximum value of the delay  $d_{\text{B5G}}$.
\begin{equation}
    \delta \geq d_{\text{MS},\text{NW}}^{send} + \text{max}\left\{d_{\text{B5G}}\right\} + d_{\text{DS},\text{SL}}^{send} +d_{\text{SL}}^{proc} 
    \label{eqn:upperbound_offset}
\end{equation}

\begin{figure}[t!]
    \centering
   \includegraphics[scale=0.149]{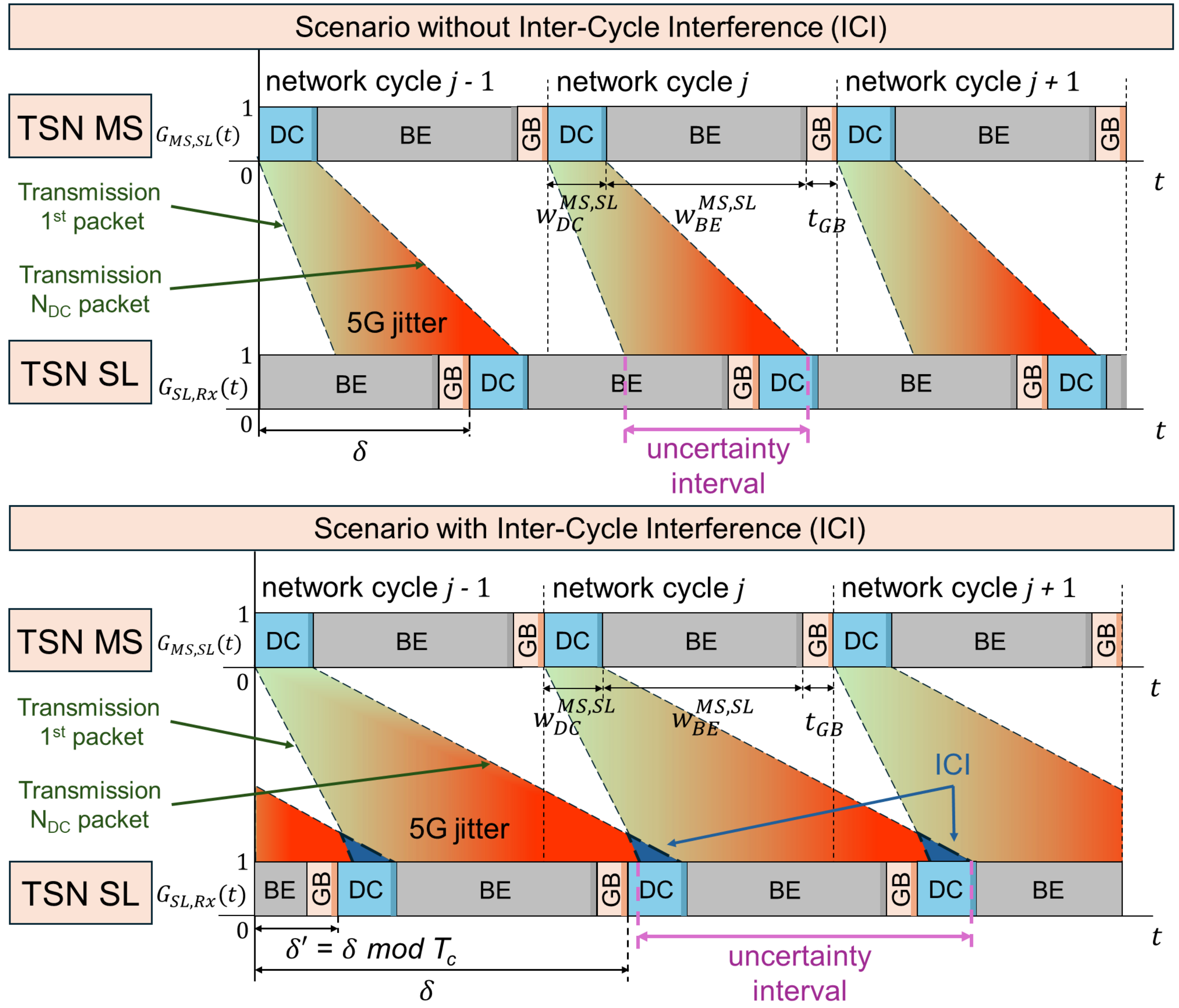}
    \caption{5G jitter effect over TAS cycles according to its uncertainty interval. Note that we assume a  node Rx $\in \mathcal{V}$ attached to TSN switch SL.}
   \label{fig:5G_jitter_effect}
\end{figure}

The offset $\delta$ must be set considering the lower bound defined in Eq.~\eqref{eqn:upperbound_offset}. In this lower bound, the only term that is variable is the propagation delay through the \gls{5G} bridge, i.e., $d_{\text{B5G}}$. In this work, we define the \textit{uncertainty interval} as the range of values which the variable $d_{\text{B5G}}$ can take. As shown in Fig. \ref{fig:5G_jitter_effect}, the impact of \gls{5G} jitter on \gls{TAS} depends on the configured \textit{transmission windows} $w_{\text{DC}}^{\text{MS},\text{SL}}$, $w_{\text{BE}}^{\text{MS},\text{SL}}$ for the \gls{TSN} switches $\text{MS}$ and $\text{SL}$;  and the \emph{network cycle} $T_{C}$. When the \textit{network cycle} exceeds the uncertainty interval, i.e., $T_{C} > d_{\text{B5G}}$, the \gls{TSN} slave can schedule their enqueued packets by simply considering the offset $\delta$ with respect to the \gls{TSN} master's scheduling. However, shorter \textit{network cycles}, i.e., $T_{C} < d_{\text{B5G}}$, and offset $\delta$ misalignments may cause packets to arrive before or after their reserved \emph{transmission windows}, leading to an increase in the packet latency $D_{tot}$. We define this phenomenon as \gls{ICI}, which can prevent packets from reaching their destinations on time. 


Since the packet transmission delay in a \gls{5G} system, i.e., $d_{\text{B5G}}$, is not strictly bounded, it is essential to consider its statistical distribution. In our work, we define the \textit{uncertainty interval} by establishing a statistical upper bound based on a specific percentile $p$ of the \gls{CDF} $F_{\text{B5G}}\left(\cdot\right)$ of this delay. Specifically, we define the upper bound $\hat{d}_{\text{B5G}}=F_{\text{B5G}}^{-1}\left(p\right)$. For example, we may set the upper bound $\hat{d}_{\text{B5G}}$ such that 99.9\% of the packets (i.e., $p=0.999$) are transmitted below this threshold. Based on that and considering Eq.~\eqref{eqn:upperbound_offset}, we define the offset $\delta$ as Eq.~\eqref{eqn:upperbound_offset2} shows. The network cycle offset $\delta'$ is calculated as  $\delta' = \delta \hspace{0.15cm}mod\hspace{0.15cm}T_{C}$.
\begin{equation}
    \delta = d_{\text{MS},\text{NW}}^{send} + \hat{d}_{\text{B5G}} + d_{\text{DS},\text{SL}}^{send} +d_{\text{SL}}^{proc} = K + \hat{d}_{\text{B5G}}
    \label{eqn:upperbound_offset2}
\end{equation}

\section{Testbed and Experimental Setup}\label{sec:Testbed}

\begin{figure*}[t!]
    \centering
   \includegraphics[scale=0.265]{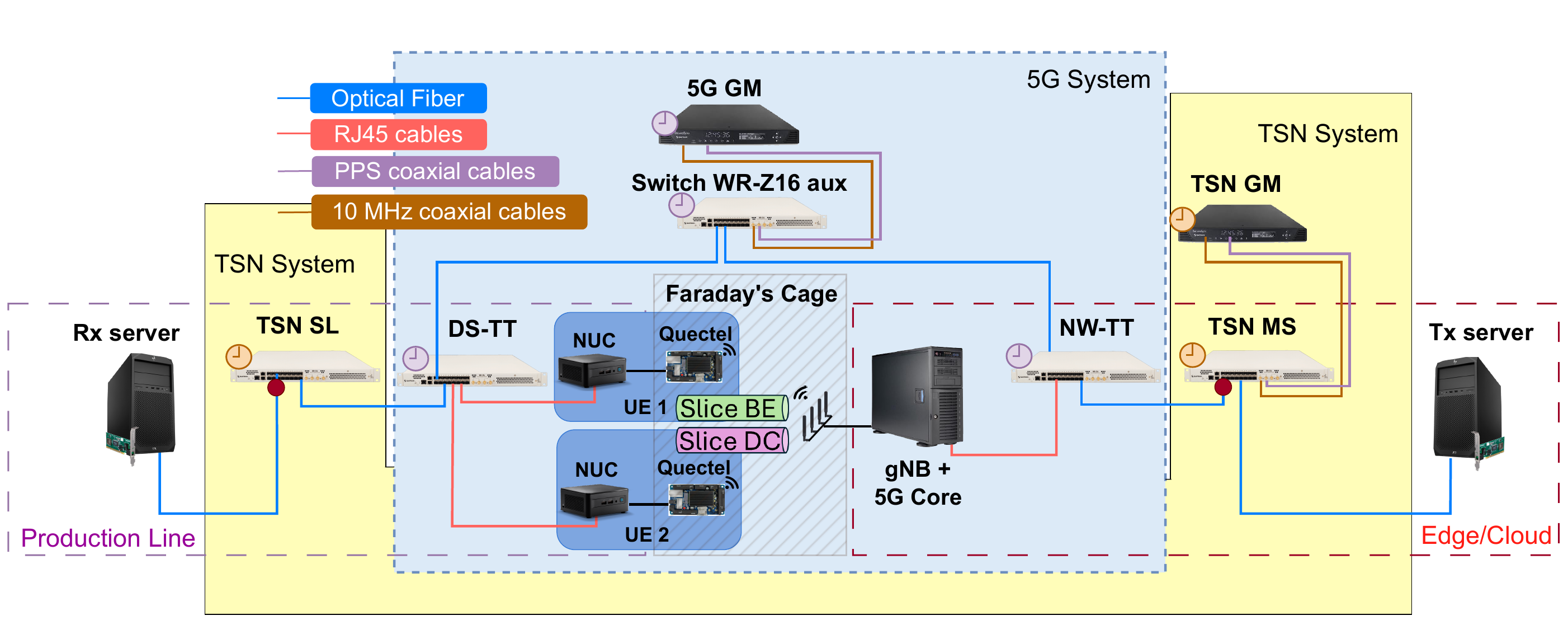}
    \caption{Proof of concept equipment and evaluated network scenario.}
   \label{fig:Testbed}
\end{figure*}

In this section, we describe the implemented \gls{5G}-\gls{TSN} testbed and the considered experimental setup.

\subsection{Testbed Description}\label{sec:testbed_description}
To validate our solution, we implemented the testbed depicted in Fig.~\ref{fig:Testbed}. It consists of: (\textit{i}) a \gls{5G} system, (\textit{ii}) five \gls{TSN} switches, (\textit{iii})  two \gls{GNSS} time synchronization servers, and (\textit{iv}) two servers acting as end devices.

\textbf{\gls{5G} System.} The \gls{5G} network consists of a single \gls{BS} and a \gls{5G} core, both built on a PC with two 50 MHz PCIe \gls{SDR} Amarisoft cards and an AMARI NW 600 license. \gls{BS} operates in the \textit{n78} frequency band, using a 30 kHz \gls{SCS} and \gls{TDD} mode. Hence, data transmission is organized into time slots. In this work, we use the 4D-2S-4U \gls{TDD} pattern, which follows a repeating pattern of four consecutive slots for \gls{DL} and four consecutive slots for \gls{UL}, with two flexible slots in between. The \gls{UE} consists of a Quectel RM500Q-GL modem connected to an Intel NUC 10 NUC10i7FNKN. The Intel NUC has a i7-10710U processor, 16 GB of RAM, and a 512 GB SSD, running Ubuntu 22.04. Two \glspl{UE} are deployed in this setup.

Experiments are conducted inside a LABIFIX Faraday cage, where the \gls{BS} antennas are connected to one of the \gls{SDR} via SMA connectors, and the Quectel modems connect via USB to the NUC thus forming the \glspl{UE}.  

The \gls{UE} does not support Ethernet-based \gls{PDU} sessions. Therefore, the \gls{5G} network is based on IP transport. Thus, to enable the transmission of industrial automation Layer 2 traffic, a \gls{VxLAN}-based tunneling mechanism is implemented. There are two different \glspl{VxLAN}, one for each \gls{PCP}-slice tuple. Furthermore, two distinct \glspl{DNN} have been configured: one for DC communications and \gls{PTP} messages, and another for BE traffic. The traffic from one \gls{DNN} is mapped to a \gls{uRLLC} slice, while the traffic from the other \gls{DNN} is mapped to an \gls{eMBB} slice. The \gls{uRLLC} and \gls{eMBB} slices are defined based on the \gls{SST} parameter of the \gls{S-NSSAI}, as specified by \gls{3GPP}, where \gls{SST}=1 corresponds to \gls{eMBB} and \gls{SST}=2 corresponds to \gls{uRLLC}.
Despite previous works consider one TSN translator per \gls{UE} and slice, we opt for a single pair of TSN translators to simplify this setup.

\textbf{TSN Network.} The \gls{TSN} switches are based on Safran’s WR-Z16 switches, with one acting as a \gls{TSN} MS, another as the \gls{TSN} SL, and two switches functioning as \gls{TSN} translators, i.e., \gls{NW-TT} and \gls{DS-TT}. The \gls{TSN} MS is directly connected to Safran’s SecureSync 2400 server to distribute a precise time reference, i.e., \gls{GM} clock, to \gls{TSN} SL. An auxiliary \gls{TSN} switch, supported by another SecureSync 2400, enables the 5G \gls{GM} clock distribution between \gls{TSN} translators.

\textbf{TSN Switch Capabilities.} WR-Z16 is based on Xilinx Zynq 7000 series \gls{FPGA} and a 1 GHz dual ARM Cortex A9 core. This allows high switching rates and very low processing delays, with configurations running on top of a Linux-based OS. Each WR-Z16 TSN switch supports IEEE 802.1Qbv \gls{TAS} and \glspl{VLAN}. It counts with 16 1 GbE \gls{SFP} configurable timing ports that can assume the role of master or slave for \gls{PTP} transmission. 
Each port counts with 4 queues. Their sizes are limited to 6.8 kB, so shorter bitrates for DC communications can be tested. 
Additionally, ports count on timestamping probes for measuring high-accuracy latencies between TSN switches. The \gls{TAS} macrotick is 16 nanoseconds. 

\textbf{Testbed Clock Synchronization.} Time synchronization between the \gls{TSN} \gls{GM} clock server and the \gls{TSN} MS switch is achieved through coaxial cables carrying the \gls{PPS} and 10 MHz signals. Similarly, the auxiliary WR-Z16 switch synchronizes with the \gls{5G} \gls{GM} clock server via coaxial cables to distribute the time reference between \gls{NW-TT} and \gls{DS-TT}. Thus, we assume very high accuracy synchronization in the 5G segment. Synchronization error between these devices is left for future research work. In our testbed, the \gls{TSN} MS and the \gls{TSN} SL use \gls{UDP} over IPv4 in unicast mode with the \gls{E2E} delay mechanism to encapsulate \gls{PTP} frames. The \gls{PTP} transmission rate is set to 1 packet per second. 

\textbf{End devices.} Two servers running Ubuntu 22.04 LTS act as a packet generator and packet sink.


\subsection{Experimental Setup}
We aim to measure the packet transmission delay in the described \gls{5G}-\gls{TSN} network for \gls{DL} direction. The delay measurement points within the considered system, i.e., $D_{emp}$ as defined in Eq.~ \eqref{eqn:empirical_delay}, are illustrated in Fig.~\ref{fig:Testbed}. Specifically, these points are the output ports of the \gls{TSN} switches MS and SL and are marked with a red dot. Network traffic is generated at the Tx server with \textit{packETH} tool. 
Packets are tagged with \gls{PCP} 2 for DC and with \gls{PCP} 0 for BE to be then matched to the \gls{uRLLC} and \gls{eMBB} \gls{5G} slices, respectively. The DC packet size is fixed to 200 Bytes and the generation rate is set to the maximum rate to allow the availability of packets at the \gls{TSN} MS queue at all times, with no impact on their discarding. In contrast, BE packets are \gls{MTU}-sized, i.e., 1500 Bytes, and generated at a constant speed of 30 Mbps. Nevertheless, due to WR-Z16's queue size limitations explained above, a maximum throughput of 1.6 Mbps for \emph{delay-citical} traffic has been fixed by window sizing for \gls{TAS} experiments to ensure no packet is discarded/lost when it finds a queue crowded. 
For each experiment, traffic was captured for 15 minutes, ensuring a minimum of 250,000 samples.

To measure packet transmission delay, we employed the WR-Z16 timestamp probes. These probes dump the sequence number allocated within the first 4 Bytes of the \gls{UDP} payload of a packet and the timestamp when it left the port to a CSV file. With the file probed both at the \gls{TSN} MS and the \gls{TSN} SL, the latencies for every packet can be calculated as the difference between them, identified by their sequence number. Note that, for accurate delay calculation, both TSN switches must be correctly synchronized, so \gls{PTP} packets also join the \gls{uRLLC} slice. 





\section{Performance Results}\label{sec:Results}
In this work, we have carried out three different experiments. 
Firstly, we have measured the \gls{5G} jitter between \gls{TSN} switches to check the effect of increments on the load. Secondly, the offset between \gls{GCL} schedules performed at \gls{TSN} switches has been swept according to the worst-case latency value. 
Finally, for this last offset configuration, the periodicity has been also swept keeping the same throughput.

\subsection{Jitter Analysis Based on Throughput}
\begin{figure}[t!]
    \centering
   \includegraphics[width=\columnwidth]{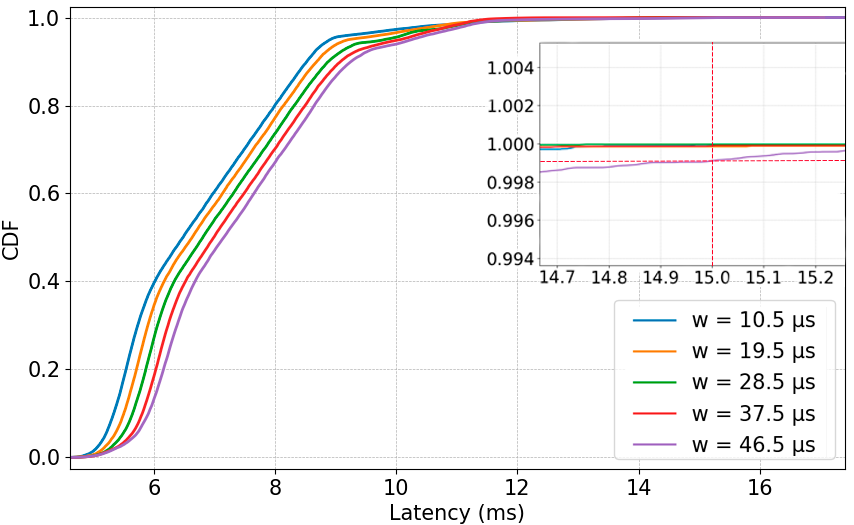}
    \caption{5G latencies CDF for different windowed loads}
   \label{fig:5G_jitter_wdw}
\end{figure}

Fig. \ref{fig:5G_jitter_wdw} presents the resulting \gls{CDF} of latency between \gls{TSN} switches MS and SL for increasing throughput via its window size, $w^{MS,SL}_{DC}$ (defined in Eq. \eqref{eqn:window_bounds}), in a TSN-enabled network incorporating a \gls{5G} bridge. The \gls{TAS} is configured at \gls{TSN} MS only to measure the $D_{emp}$ delay (Fig. \ref{fig:network_model}). The \emph{network cycle} is fixed at 30 ms. A sweep in throughput steps of 300 kbps, from 350 kbps to 1.55 Mbps, results in window sizes $w^{MS,SL}_{DC}$ of 10.5 $\mu$s to 46.5 $\mu$s, respectively. It has been proven that configurations with smaller window sizes exhibit the lowest latencies, whereas larger windows introduce a slight rightward delay shift as $w^{MS,SL}_{DC}$ increases. This behavior suggests that increasing this window substantially affects scheduling efficiency, potentially due to buffering mechanisms within the \gls{5G} bridge. This effect takes place for each DC packet while packets ahead in the queue are still awaiting their transmission, as shown with Eq. (\ref{eqn:empirical_delay}). The convergence of all distributions suggests averages between 6.5 ms and 7.25 ms but a much higher maximum value at 17 ms. However, we conclude with a well-defined latency upper bound near 15 ms for that 99.9\% reliability as commented in Section \ref{sec:jitter}. This value provides the empirical offset $\delta$ to be applied between MS and SL.

\subsection{Jitter Analysis Based on Offset Between TSN MS and SL}
\begin{figure}[b!]
    \centering
   \includegraphics[width=\columnwidth]{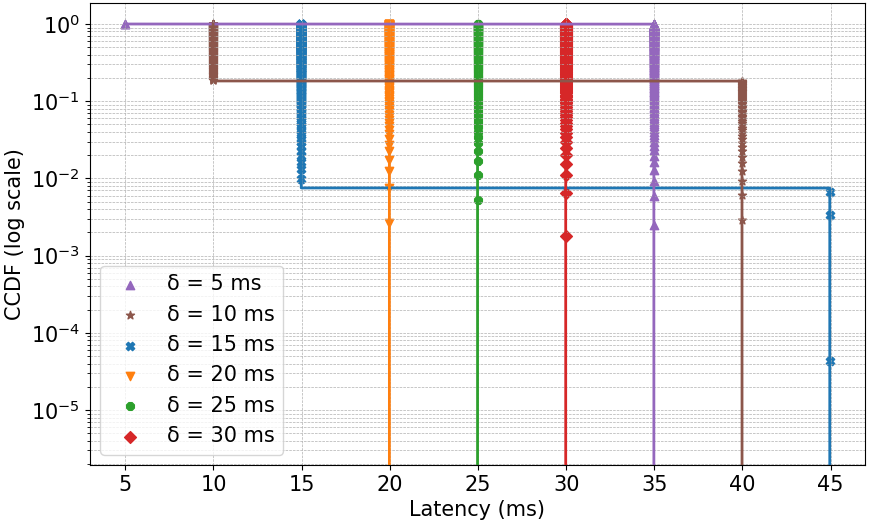}
    \caption{Offset sweep for fixed period and window values}
   \label{fig:test_offset}
\end{figure}

This second experiment has focused on the highest latency configuration to characterize the value of the offset $\delta$ that best fits to optimize latency for DC traffic. The time duration of the window $w^{MS,SL}_{DC}$ and the period of the \gls{TSN} MS switch are kept constant. Thus, the size of $w^{MS,SL}_{DC}$ is 46.5 $\mu$s, and the \emph{network cycle} $T_{C}$ is 30 ms for both \gls{TSN} MS and SL. Then, the configuration of the \gls{TSN} slave switch is identical, except that we sweep the offset $\delta$ over the range [5, 30] ms. The step is 5 ms. Fig. \ref{fig:test_offset} shows the \gls{CCDF} of the results, where latencies decrease as its offset gets closer to the worst-case latency measured in the previous experiment, i.e., 15 ms. However, at this point, latencies start to jump the equivalent of $T_C$, i.e., 30 ms, as packets now have to wait until the next window for DC traffic in the following \emph{network cycle}. This means that \gls{ICI} cumulative effect is occurring in the queue, the same as seen in Section \ref{sec:jitter}. Note that the effect of packet accumulation for $\delta = $ 15 ms arises at this point as this jump probability is around 0.7\%, higher than the 0.1\% objective set in the previous experiment. The lower the offset, the higher the probability of this jump. For $\delta =$ 5 ms, the probability of this latency jump is over 90\% as this is too tight. A lower value of offset increases this probability until almost every packet arrives to \gls{TSN} slave later than expected. 
On the other hand, the offset value of $\delta =$ 20 ms optimizes the latency without a deeper sweep. 

\subsection{Jitter Analysis Based on Network Cycle}
\begin{figure}[t!]
    \centering
   \includegraphics[width=\columnwidth]{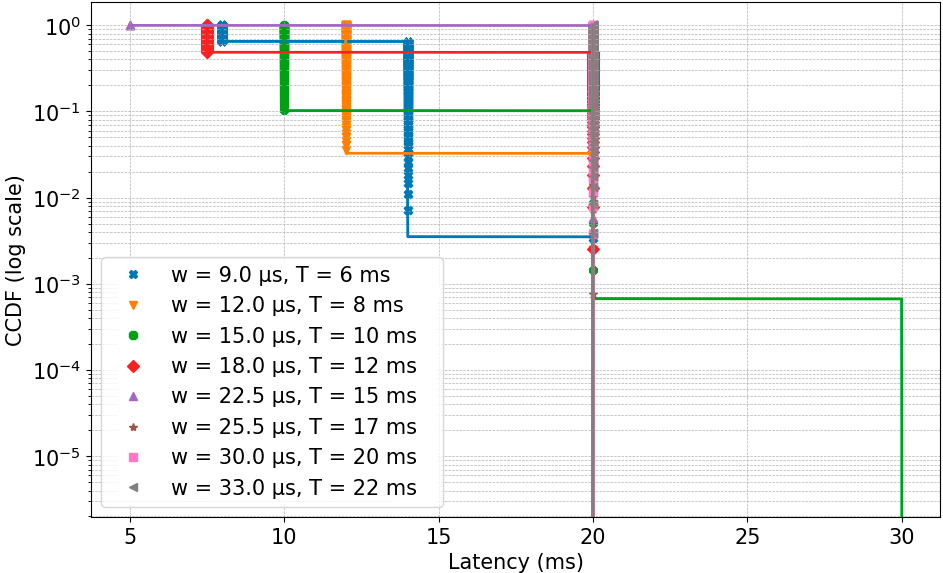}
    \caption{Window and period sweep for fixed offset}
   \label{fig:test_2}
\end{figure}

In this third and last experiment, we used the same setup as in the previous experiment, i.e., $\delta =$ 20 ms, but we now sweep the \emph{network cycle} $T_C$ instead of the offset $\delta$. To maintain the throughput, we calculate a proportional window $w^{SL,Rx}_{DC}$ at the \gls{TSN} SL for each $T_C$ value. As a contingency measure, we applied an additional 25\% to $w^{SL,Rx}_{DC}$ displayed in Fig. \ref{fig:test_2} at the \gls{TSN} SL to avoid overloading the queue at the expense of slightly reducing bandwidth for the BE traffic. Decreasing $T_C$, and therefore increasing the frequency of \emph{transmission window} $w^{SL,Rx}_{DC}$, results in lower latencies than the fixed offset $\delta$, aligned with the corresponding \gls{CDF}. However, this may limit the total bandwidth for BE traffic as the number of windows and the \emph{guard bands} increase per unit of time. Additionally, this may cause the \gls{ICI} effect (Section \ref{sec:jitter}), and packets will start to accumulate between network cycles until the queue is saturated. In our results, $T_C$ values below 15 ms start suffering \gls{ICI}, in which lower latencies can be seen like the case of 7.25 ms for $T_C=$ 12 ms. However, these probabilities are very low, and the corresponding latencies increase rapidly. For $T_C=$ 6 ms and $T_C=$ 10 ms, the latency increases to a third window at 20 ms and 30 ms, respectively; which implies some packets are retained in queue for two network cycles. In the case of $T_C=$ 8 ms this effect does not occur. The lowest latency value for each $T_C$ depends on its alignment with the offset $\delta$. Thus, for the same traffic load, \gls{TAS} modifies the traffic pattern in \gls{5G} and, therefore, the latencies of the \gls{5G}-\gls{TSN} network.

\section{Conclusions and Future Work}\label{sec:Conclusions}
In this work, we presented an empirical analysis of the impact of 5G jitter on the performance of IEEE 802.1Qbv in an integrated \gls{5G}-\gls{TSN} network. Firstly, we have provided background on aspects of industrial \gls{5G}-\gls{TSN} networks, especially those related to elements that compose this kind of network and its applications. Secondly, we have analyzed, formulated, and modeled the effect of 5G jitter on the \gls{TAS} mechanisms through the associated constraints. We have focused this on the \gls{DL} communication between two \gls{TSN} switches interconnected through a \gls{5G} bridge and transferred to our testbed equipment, including a set of \gls{TSN} translators. Finally, we have identified and tested the key configuration parameters required to optimize the \gls{TAS} scheduling for this integration in a 5G-TSN environment. Our experiments show that the offset between \gls{TSN} switches plays a key role in achieving determinism, beyond the latencies measured in the \gls{5G} system. 

Our line of future work is intended to perform additional experiments related to the periodicity and the \gls{ICI} effects to optimize traffic scheduling for general applications.

\bibliographystyle{ieeetr}
\bibliography{references}

\end{document}